\documentclass[journal]{IEEEtran}
\usepackage{array}
\usepackage{graphicx}
\usepackage[cmex10]{amsmath}
\usepackage{subfigure}
\usepackage{floatrow}
\usepackage[font = footnotesize]{caption}
\usepackage{cite}

\begin{document}
\title{Low-power Spin Valve Logic using Spin-transfer Torque with Automotion of Domain Walls}
\author{Sou-Chi~Chang, Sasikanth~Manipatruni,~\IEEEmembership{Member,~IEEE}, Dmitri~E.~Nikonov,~\IEEEmembership{Senior~Member,~IEEE}, Ian~A.~Young,~\IEEEmembership{Fellow,~IEEE}, and Azad~Naeemi,~\IEEEmembership{Senior~Member,~IEEE}
\thanks{S. -C. Chang and A. Naeemi are with School of Electrical and Computer Engineering, Georgia Institute of Technology, Atlanta, GA 30332 USA.}
\thanks{S. Manipatruni, D. E. Nikonov, and I. A. Young are with Components Research, Intel Corporation, Hillsboro, OR 97124 USA. (email: souchi@gatech.edu)}}%
\markboth{Draft Copy}
{Shell \MakeLowercase{\textit{et al.}}}

\maketitle
\begin{abstract}
A novel scheme for non-volatile digital computation is proposed using spin-transfer torque (STT) and automotion of magnetic domain walls (DWs). The basic computing element is composed of a lateral spin valve (SV) with two ferromagnetic (FM) wires served as interconnects, where DW automotion is used to propagate the information from one device to another. The non-reciprocity of both device and interconnect is realized by sizing different contact areas at the input and the output as well as enhancing the local damping mechanism. The proposed logic is suitable for scaling due to a high energy barrier provided by a long FM wire. Compared to the scheme based on non-local spin valves (NLSVs) in the previous proposal, the devices can be operated at lower current density due to utilizing all injected spins for local magnetization reversals, and thus improve both energy efficiency and resistance to electromigration. This device concept is justified by simulating a buffer, an inverter, and a 3-input majority gate with comprehensive numerical simulations, including spin transport through the FM/non-magnetic (NM) interfaces as well as the NM channel and stochastic magnetization dynamics inside FM wires. In addition to digital computing, the proposed framework can also be used as a transducer between DWs and spin currents for higher wiring flexibility in the interconnect network.
\end{abstract}

Keywords - spin-transfer torque, domain wall, digital logic

\section{Introduction}
Spintronics, a field of switching magnetization using variety of sources \cite{PhysRevApplied.4.047001}, has recently been one of the most promising candidates in the beyond complementary metal-oxide-semiconductor (CMOS) computing \cite{7076743}, as power dissipation due to leakage currents in present-day integrated circuits  increases with device density, doubling approximately every two years according to the Moore's law \cite{Moore1965}. The major advantage of encoding digital information into magnetic states is their non-volatility, which eliminates the delay and energy required to save and fetch the data when a microprocessor is put in a $\it{sleep}$ state, with power off. It thus loosens the power constraints in a microprocessor. In most of the proposed spin-based logic devices \cite{7076743}, the bit is represented by the magnetization of a single-domain ferromagnet, and the communication between bits is realized using either spin currents \cite{asl,Zutic2005}, the dipolar coupling \cite{4769989,Cowburn25022000}, or spin wave propagation between magnetoelectric cells \cite{:/content/aip/journal/jap/106/12/10.1063/1.3267152}. However, no matter how well the data is preserved while bits are transmitted, the data retention time is degraded as the device size is scaled due to the lowering of the energy barrier of the magnet \cite{PhysRevB.62.570}. Hence, it is of interest to seek an alternative magnetic structure in digital computing, where bits can be retained longer in the path of scaling.

\begin{figure}[htp]
\includegraphics[width = 3.5in]{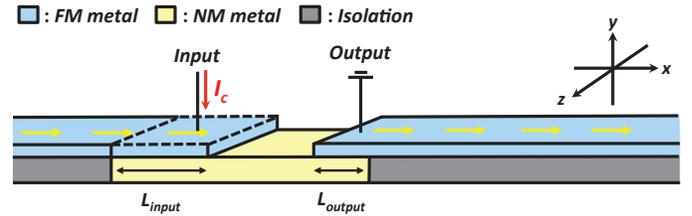}
\caption{The schematic of the proposed computing element, comprising a lateral metallic SV and two FM interconnects, where DW automotion is used to update the magnetization of the wire. The damping mechanism in the region bounded by dashed lines is stronger than that in the rest to enhance the logic non-reciprocity, which is also optimized by sizing the input and output contact areas (or varying $L_{input}$ and $L_{output}$). The blue, gray, and yellow colors designate FM, insulating, and NM materials, respectively. The yellow arrow represents the magnetization orientation in the wire. The magnetizations pointing to $+x$ and $-x$ are defined as $1$ and $0$, respectively.}
\label{fig1}
\end{figure}

Using FM nanowires to store bits in digital logic can provide better non-volatility than using a single-domain magnet due to a higher energy barrier of the wire. Some spin-based devices have been proposed to realize computation by controlling the location of the magnetic DW in the FM wire using in-plane spin currents \cite{6177298,6332558}. However, those devices need highly resistive magnetic tunneling junctions (MTJs) to convert the magnetic signal to the electrical one. The latter is used to drive magnetization switching in the next stage for a more complex logic function. On the other hand, the general concept of using FM wires as interconnects has been proposed, in which the shape-anisotropy-driven DW motion, also known as DW automotion \cite{PhysRevB.82.214414}, is used to update the bit inside the wire \cite{:/content/aip/journal/jap/115/21/10.1063/1.4881061}. Recently, devices in the form of NLSVs connected by FM wires based on automotion of DW has been proposed due to full metallic  structures and the possibility of energy-free propagation of DW once it is created \cite{7130566}. In contrast, in NLSVs, only part of injected spins contribute STT to DW creation in the interconnect, and STT becomes much weaker as the shunt path in the device is reduced \cite{7384517}. As a result, a lateral simple spin-valve with a tunneling barrier is suggested to eliminate the shunt path and simultaneously maintain the non-reciprocity \cite{6834821,6818391}. While the device is able to fully use injected spins for information processing, it still has some drawbacks due to the fact that a tunneling barrier, that is required to maintain the non-reciprocity \cite{5966345} or overcome conductivity mismatch \cite{PhysRevB.62.R16267}, is highly resistive. As a result, in this paper, to further reduce the device resistance while all the injected spins are used to manipulate the magnetization, a metallic SV with FM interconnects based on DW automotion is proposed as a basic element for digital computing (Fig. \ref{fig1}). Unlike typical applications such as magnetic field sensors and random access memories (RAM), where the asymmetry of SVs is achieved by either making one magnet thicker than another or having the exchange bias provided by an antiferromagnet on one of the magnets, here the logic non-reciprocity is realized by sizing different contact areas of the input and the output as well as locally enhancing the damping process underneath the input contact.

The rest of this paper is organized as follows: In Section II, physics underlying this new computation scheme and the method to implement an inverter, a buffer, and a $3$-input majority gate are presented. In Section III, a theoretical framework of numerically modeling both device and interconnect is presented in detail. In Section IV, simulations based on the numerical model are shown to justify the proposed concept, and some limitations and design aspects are discussed. Section V concludes the paper.
\section{Device Physics}
This section covers essential physics behind the proposed device and also shows how to implement functions such as an inverter, a buffer, and a majority gate under this scheme.

Figure \ref{fig1} shows the device structure, comprising a simple metallic SV with two FM wires as interconnects. A voltage is applied across the metallic channel. Since the resistance of the FM wire is much higher than that of the NM channel thanks to its small thickness, almost all of the current flows into the NM channel rather than the FM wire. When a negative voltage is applied at the input, electrons with spin polarization collinear to the magnetization underneath the input contact are injected into the NM channel. Because of strong screening effects inside the normal metal, injected spins travel through the channel mainly by diffusion \cite{PhysRevB.48.7099}. As spin-polarized electrons reach the end of the metallic channel, STT is exerted onto the FM region under the output contact. As long as the magnitude of STT is well above the threshold for local magnetization reversal and within a certain range, a DW with a $+x$ velocity can be created in the beginning of the wire \cite{:/content/aip/journal/jap/115/21/10.1063/1.4881061}. The DW would propagate toward the end of the FM wire due to intrinsic shape anisotropy \cite{PhysRevB.82.214414}. Therefore, the bit is written from the input to the output and passed to the next stage through the interconnect. Furthermore, at the output, while a DW is generated by the input, electrons with spin polarization anti-parallel to the magnetization underneath the output contact are accumulated at the interface and diffuse back to the input. Thus, there is also STT exerted on the FM region at the input. To reduce STT effects on the input for non-reciprocity, the damping mechanism at the input is set to be stronger than the rest of the interconnect, making the input's response to STT weaker. Note that local highly-damped regions may be realized by intentionally increasing the impurity concentration (e.g. Nd) at the end of the FM wire \cite{PhysRevB.89.184412}. Making the input highly damped can also ensure that the DW can disappear at the end of the wire and thus no data reflection occurs in the interconnect \cite{7130566}. In addition to the local enhancement of damping coefficient, the non-reciprocity can also be further improved by varying charge current density at both input and output, which strongly influence the magnitude of STT in the device (see Supplementary Materials for analytical derivations). On the other hand, if the voltage polarity is reversed, STT experienced by the output is due to back-diffusive spin-polarized electrons from the input interface. Since spin polarization of these electrons is anti-parallel to the input magnetization, the bit is inverted as it is written into the output.

\begin{figure}[htp]
\includegraphics[width = 3.6in]{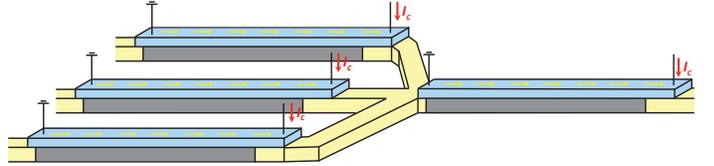}
\caption{The schematic of a three-inputs majority gate under the proposed scheme. An AND or an OR gate can be realized by setting one of the inputs as a control terminal.}
\label{fig2}
\end{figure}

To sum up, the device is switched to lower-energy parallel and anti-parallel configurations as negative and positive voltages are applied, respectively. This feature is of interest in digital computing since as the bit is encoded into the magnetization, and a non-inverting or inverting logic can be realized using the same structure by simply changing the supply voltage polarity. The device operation can be understood by making an analogy to a simple spin valve with one layer fixed and another layer free. Here a "quasi-fixed" layer is achieved by both reducing the STT applied at the input and the input's response to STT. In addition, a 3-input majority gate is also of interest since majority logic is more efficient for implementing combinatorial logic (fewer gates required). An AND or an OR gate can be realized by setting one of the inputs as a control terminal \cite{5928383}. Figure \ref{fig2} shows a 3-input majority gate implementation using the proposed scheme. Similar to Ref. \cite{5928383}, the magnetization underneath the output contact is controlled by the net spin polarization of current flowing into the output, which is mainly determined by the majority of the inputs. Note that for a 3-input majority gate, if magnetizations of the inputs are not identical (e.g. two inputs are in the $+x$ direction, and one input is in the $-x$ direction), net STT at the output is not simply the sum over all the contributions from the inputs. In other words, the net spin torque is weakened under non-identical inputs because there are some currents directly flowing from one input to another due to different resistances of the signal paths (e.g. parallel and anti-parallel configurations result in low and high resistances, respectively).

\section{Mathematical Models}
To model the proposed scheme as shown in Fig. \ref{fig1}, spin circuit theory is used to describe spin transport through NM and FM metals as well as their interfaces \cite{Brataas2006157,6359950,Camsari2015}. Note that in this work, at the NM/FM interface, the perpendicular spin component is simply dependent on spin accumulation at the NM side \cite{PhysRevB.66.224424,Camsari2015}, instead of that across FM/NM interface assumed in Ref. \cite{6359950}. For FM nanowires, the magnetization dynamics is captured by the stochastic Landau-Lifshitz-Gilbert (LLG) equation. Similar to Ref. \cite{7130566}, a self-consistent numerical iteration between spin circuits and the LLG equation is required to describe the time evolution of the system. In this section, the theoretical approach to model the proposed scheme is presented in detail.

\begin{figure}[htp]
\includegraphics[width = 3.6in]{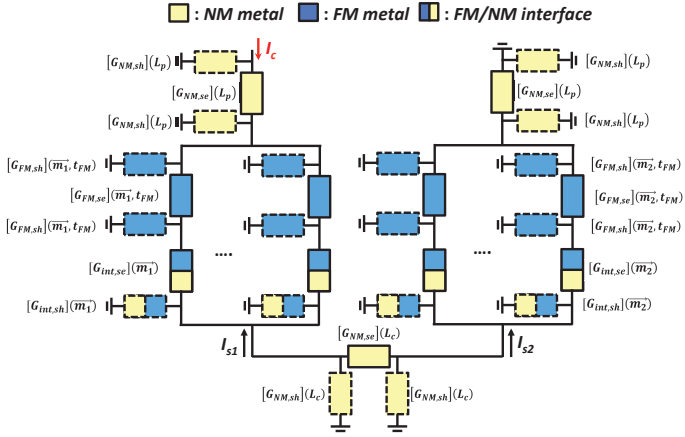}
\caption{An equivalent spin circuit of a single device shown in Fig. \ref{fig1}. Blue, yellow, and mixed blue/yellow bars stand for FM materials, NM materials, and FM/NM interfaces, respectively. Black and red arrows stand for spin currents flowing into the FM materials and charge current sources, respectively. Dashed and solid bars represent shunt and series conductances, respectively.}
\label{fig3}
\end{figure}

An equivalent circuit of a single device is given in Fig. \ref{fig3}, where $\left[G_{NM,se}\right]$ and $\left[G_{NM,sh}\right]$ are the series and shunt conductances of the NM material, respectively, $\left[G_{FM,se}\right]$ and $\left[G_{FM,sh}\right]$ are the series and shunt conductances of the FM material, respectively, $\left[G_{int,se}\right]$ and $\left[G_{int,sh}\right]$ are series and shunt conductances at the FM/NM interfaces, respectively, $L_{p}$, $L_{c}$, and $t_{FM}$ are the contact length, channel length, and thickness of FM wires, respectively, and $\vec{m_{1}}$ as well as $\vec{m_{2}}$ are the magentizations at the input and output, respectively. Note that all the conductances in Fig. \ref{fig3} are 4$\times$4 matrices, and $\left[G_{int,se}\right]$, $\left[G_{int,sh}\right]$, $\left[G_{FM,se}\right]$, as well as $\left[G_{FM,sh}\right]$ vary with the local magnetization and thus is space-dependent. The number of FM and interface sub-circuits at both input and output is determined by the size of the contacts (i.e. $L_{input}$ and $L_{output}$ in Fig. \ref{fig1}). The detailed mathematical expressions of these conductance matrices can be found in Refs. \cite{6359950,Camsari2015}. In this work, the interface shunt conductance is only required at the side where STT is present \cite{Camsari2015}, rather than both sides assumed in Refs. \cite{6547730,Srinivasan2014}. This is because from both experimental and theoretical studies, the transverse spin component can only penetrate some typical FM metals (e.g. Fe, Co, and Py) less than $1$nm \cite{PhysRevB.71.064420,PhysRevB.66.014407,PhysRevB.71.100401}. However, in some weak FM metals such as CuNi, the transverse spin component may appear at both sides of the FM thin film, and thus a more sophisticated expression for the interface transport is required \cite{PhysRevB.73.054407}. The overall conductance matrix for a single device, $\left[G\right]_{4N\times 4N}$, can be obtained using a nodal analysis similar to Ref. \cite{7130566}, where a set of equations are established by the fact that the net current at each node is zero and satisfies the following equation:
\begin{eqnarray}
\left[I\right]_{4N\times 1}=\left[G\right]_{4N\times 4N}\left[V\right]_{4N\times 1},
\label{eq1}
\end{eqnarray}
where $\left[I\right]_{4N\times 1}$ and $\left[V\right]_{4N\times 1}$ are current and voltage column vectors describing charge and spin components for all the nodes in the circuit, $N$ is the total number of nodes in the circuit, and the index $4$ includes one charge component and three spin elements in the x, y, and z directions. By multiplying the inverse matrix of $\left[G\right]_{4N\times 4N}$ at both sides of Eq. \ref{eq1}, the charge and spin voltages at each node can be obtained. And the nodal current vector, $\left[I_{ij}\right]$, including both charge and spin components flowing from the $i$ to the $j$ node, is given as
\begin{eqnarray}
\left[I_{ij}\right]_{4\times 1}=\left[G_{ij}\right]_{4\times 4}\left(\left[V_{i}\right]_{4\times 1}-\left[V_{j}\right]_{4\times 1}\right),
\label{eq2}
\end{eqnarray}
where $\left[G_{ij}\right]_{4\times 4}$ is the conductance with $\left[I_{ij}\right]_{4\times 1}$ flowing through, and $\left[V\right]_{4\times 1}$ is the nodal voltage vector having both charge and spin components. Thus, spin currents flowing into the input and the output of a single device can be calculated by solving the nodal equations of the equivalent circuit shown in Fig. \ref{fig3}. To evaluate the magnetic responses of the wires, spin currents flowing into the FM wires at both input and output, calculated from the spin circuit, become the inputs to the stochastic LLG equation given as
\begin{eqnarray}
\frac{\partial \vec{m}}{\partial t}&=&-\gamma \mu_{0}\left(\vec{m}\times\vec{H_{eff}}\right)+\alpha\left(\vec{m}\times\frac{\partial \vec{m}}{\partial t}\right) \nonumber \\
& & +\frac{\vec{m}\times\left(\vec{I_{s}}\times\vec{m}\right)}{eN_{s}},
\label{eq3}
\end{eqnarray}
where $\vec{m}$ is the unit vector in the direction of magnetization, $\gamma$ is the gyromagnetic ratio, $\mu_{0}$ is the free space permeability, $\alpha$ is the Gilbert damping coefficient, $I_{s}$ is the average spin current flowing into the FM wires, $e$ is the magnitude of the electron charge, and $N_{s}$ is the number of Bohr magnetons inside the magnet, defined as $\frac{2M_{s}V_{FM}}{\gamma \hbar}$ with $V_{FM}$ being the volume of the FM wire in contact to the NM channel, $M_{s}$ being the saturation magnetization, and $\hbar$ being the reduced Planck constant. Note that $\vec{H_{eff}}$ in Eq. \ref{eq3} is the effective magnetic field including material anisotropy, $\vec{H_{m}}$, shape anisotropy, $\vec{H_{s}}$, exchange interaction, and thermal random noise, $\vec{H_{th}}$, and given as follows:
\begin{eqnarray}
\vec{H_{eff}}=\vec{H_{m}}+\vec{H_{s}}+\frac{2A}{\mu_{0}M_{s}}\frac{\partial^{2}\vec{m}}{\partial x^{2}}+\vec{H_{th}},
\label{eq5}
\end{eqnarray}
where all the internal fields are defined the same as those in Ref. \cite{7130566}, and $A$ is the exchange constant. Note that since the width and the thickness of FM wires are quite small, magentizations in both $y$ and $z$ directions are assumed to be uniform and thus the exchange field is only dependent on the $x$ direction. From Refs. \cite{:/content/aip/journal/jap/115/21/10.1063/1.4881061,7130566}, it has been shown that this assumption describes DW automotion well in terms of some important quantities of interconnect such as DW velocity and driving currents for DW creation compared to full micromagnetic simulations. Once the magnetization of the wire is updated after solving the LLG equation, the FM and FM/NM interface conductance matrices will also be changed accordingly due to their dependence on the local magnetizations. Hence, the new spin circuit has to be solved iteratively to obtain the updated spin currents flowing into the wire, and a self-consistent numerical solution between the spin circuit and the LLG equation establishes a complete dynamics of a single device. Similarly, a 3-input majority gate can be simulated using the equivalent circuit shown in the Supplementary Materials with the LLG equations for FM wires.
\section{Results and Discussion}
In this section, the proposed concept is justified using the numerical scheme mentioned in the preceding section by investigating the logic non-reciprocity, and a buffer, an inverter, as well as a 3-input majority gate are simulated. To explore the potential of the proposed device, its performance is also compared with NLSV and CMOS counterparts. In the following simulations, the in-plane magnetized wire is used due to a faster DW velocity compared to the out-of-plane one \cite{:/content/aip/journal/jap/115/21/10.1063/1.4881061}. The wire length is chosen as $300$nm for typical local interconnects \cite{7076743}, and the mesh size is $2$nm. The magnetic wires are modeled with the material parameters of permalloy, in which material anisotropy is weak and thus the energy barrier is mainly determined by shape anisotropy. Furthermore, copper (Cu) is used as the non-magnetic material and the length is chosen as $70$nm, which can be further reduced to improve the energy efficiency as long as the dipole coupling between the input and the output FM wires is weak enough \cite{7384517}. Note that Cu transport parameters such as resistivity and spin relaxation length are obtained through the compact model developed in Ref. \cite{6627977} by assuming that the specularity ($p$) and reflectivity ($R$) of the wire are $1$ and $0.1$, respectively. If not stated otherwise, for simplicity, the applied current pulse is set to be $1.5$ns to drive a single device. However, thanks to DW automotion, the current can be turned off to save energy immediately after the DW is created in the beginning of the wire. Simulations parameters not mentioned above are summarized in Table. \ref{tab1}.

\begin{table}[ht!]
\centering
\caption{Simulation parameters in Section IV. $\rho$ and $\beta$ are the resistivity and spin polarization of conductivity for permalloy, respectively. $l_{sf,\parallel}$ and $l_{sf,\perp}$ are longitudinal and transverse spin relaxation lengths of permalloy, respectively. $l_{FM}$, $w_{FM}$, and $t_{FM}$ are the length, width, and thickness of FM wires, respectively. $l_{NM}$, $w_{NM}$, and $t_{NM}$ are the length, width, and thickness of the NM channel, respectively. $G_{\uparrow\uparrow}$, $G_{\downarrow\downarrow}$, and $G_{\uparrow\downarrow}$ are majority, minority, and mixing interface conductances, respectively.}
\renewcommand{\arraystretch}{1.2}
\begin{tabular}{c c c}
\hline
\hline
Symbol & Value & Unit \\
\hline
$\rho$ & $1.4\times 10 ^{-7}$ \cite{:/content/aip/journal/jap/45/6/10.1063/1.1663668} & $\Omega\cdot$m \\
$\beta$ & $0.6$ \cite{Brataas2006157} & - \\
$l_{sf,\parallel}$, $l_{sf,\perp}$ & $5$ \cite{Bass1999274}, $0.8$ \cite{PhysRevB.71.100401} & nm \\
$M_{s}$ & $8\times10^{5}$ \cite{oommf} & A$\cdot$m$^{-1}$ \\
$A$ & $1.3\times 10^{-11}$ \cite{oommf} & joule$\cdot$m$^{-1}$ \\
$\alpha$ & $0.007$ \cite{PhysRevB.92.140413} & - \\
$l_{FM}$, $w_{FM}$, $t_{FM}$ & $300$, $20$, $2$ & nm \\
$l_{NM}$, $w_{NM}$, $t_{NM}$ & $70$, $20$, $20$ & nm \\
$G_{\uparrow\uparrow}$, $G_{\downarrow\downarrow}$, $G_{\uparrow\downarrow}$ & $0.9$, $0.1$, $0.39$ \cite{Brataas2006157} & $10^{15}\Omega^{-1}\cdot$m$^{-2}$ \\
\hline
\hline
\end{tabular}
\label{tab1}
\end{table}

\begin{figure}[htp]
\includegraphics[width = 3.5in]{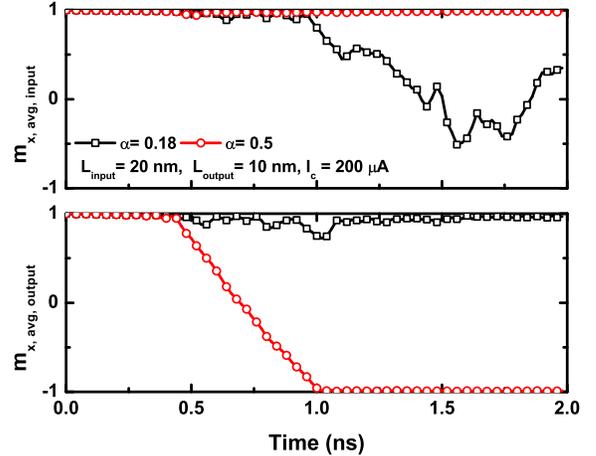}
\caption{Time evolution of average magnetizations of FM wires for the input (top) and the output (bottom) in a SV inverter with different damping coefficients at the end of the wire.}
\label{fig4}
\end{figure}

\subsection{Logic Non-reciprocity}
To maintain the device non-reciprocity, it is of importance to ensure that only the input can affect the output, not the other way around. Hence, in our proposed scheme based on SVs, the non-reciprocity is created as the magnetic response to STT of the input is much weaker than that of the output, which is realized by increasing the damping process at the end of the interconnect. Figure \ref{fig4} shows the effect of damping coefficient at the end of the FM wire on the non-reciprocity as both input and output experience strong STT. In Fig. \ref{fig4}, it can be seen that in the case of the damping coefficient being $0.18$, the magnetization at the input is quite sensitive to STT, and thus an inverter cannot be operated normally under positive bias current. However, as damping coefficient is increased to $0.5$, the input's magnetization is almost unperturbed even under strong STT, and in such a case, the output magnetization can be switched as expected. Note that the reversal of the average magnetization in an FM wire implies the fact that a DW is created by STT in the beginning of the wire, travels automatically using intrinsic shape anisotropy through the channel, and disappears at the end of the wire through the damping process. Although large damping coefficient is desirable in the proposed scheme, so far the highest one demonstrated experimentally is only about $0.18$ by intentionally doping Pd into Py \cite{PhysRevB.89.184412}. Therefore, another efficient way to improve the non-reciprocity is to reduce STT exerted on the input by sizing the contact areas of both input and output. However, since the input and the output are closely coupled in a SV, it is impossible to fix the STT at the output while reducing that at the input. As a result, properly choosing the input and output contact areas is critical for the non-reciprocity especially when the damping coefficient at the input is fixed. 

\begin{figure}[htp]
\includegraphics[width = 3.5in]{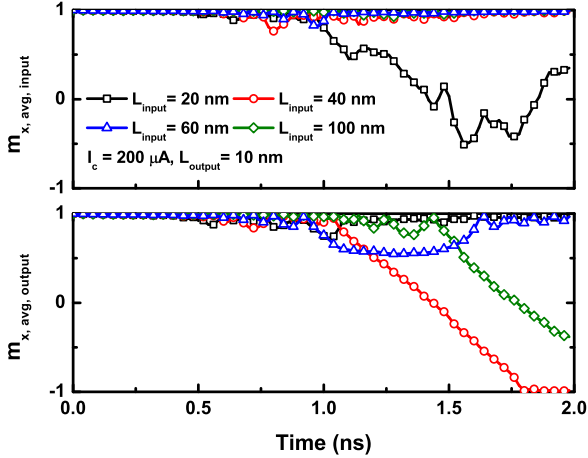}
\caption{Time evolution of average magnetizations of FM wires for the input (top) and the output (bottom) in a SV inverter with different input contact lengths. The damping coefficient at the end of FM interconnect is set to be $0.18$ for Figs. \ref{fig5} to \ref{fig13}.}
\label{fig5}
\end{figure}

In Fig. \ref{fig5}, as the input contact length is increased from $20$nm (black) to $40$nm (red), the input's magnetization becomes less disturbed due to weaker STT resulting from smaller current density flowing through the input. However, since the magnetization at the input is still affected by STT due to the output, one can observe that input's and output's magnetizations still interact strongly with each other before the DW at the output is created. The input can become less sensitive to STT by further increasing the input contact length to $60$nm (blue). In such a case, the coupling between the input and the output is significantly reduced, and a DW can be created faster. Note that there is no guarantee that a DW can definitely reach the end after creation, since a DW with $-x$ velocity may be created and then disappears in the beginning of the wire (e.g. blue in Fig. \ref{fig5}). If $100$nm is used as the input contact length (olive), the input's magnetization is almost insensitive to STT, similar to setting the damping coefficient as $0.5$ as shown in Fig. \ref{fig4}. However, the time required to create a DW becomes longer because STT at the output is weakened as the charge current density at the input becomes too smaller.

\begin{figure}[htp]
\includegraphics[width = 3.5in]{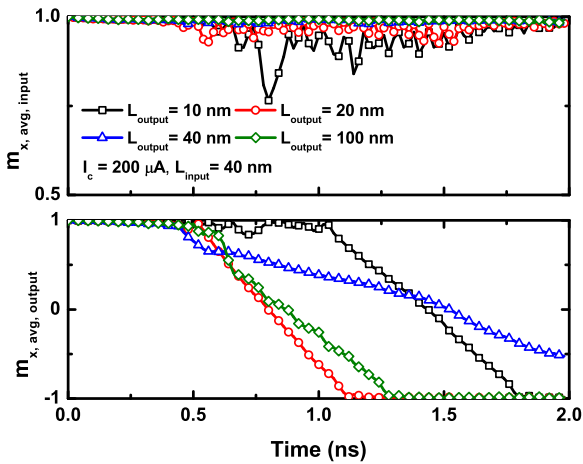}
\caption{Time evolution of average magnetizations of FM wires for the input (top) and the output (bottom) in a SV inverter with different output contact lengths.}
\label{fig6}
\end{figure}

Since in SVs, the input and the output are closely coupled through STT, the sensitivity of the input to STT can also be reduced by increasing the output contact length. In Fig. \ref{fig6}, it is shown that by increasing the output contact length from $10$nm (black) to $20$nm (red), the time that a DW is created becomes faster. This is because STT at the input is also reduced by lower current density at the output. The output switching becomes more efficient as the input is less sensitive to STT even though the output current density is reduced.  As the length increases to $40$nm (blue), the input's magnetization is almost unaffected by the output and thus an improvement in the speed of DW creation is also observed. However, as discussed in Ref. \cite{7130566}, a DW that is created faster may not reach the end of the wire earlier because of its slow DW velocity (e.g. blue in Fig. \ref{fig6}). The speed of DW creation is not further improved as the output contact length is increased to $100$nm because the input's magnetization is already stable enough for efficient switching at smaller output contact lengths.

\begin{figure}[htp]
\includegraphics[width = 3.5in]{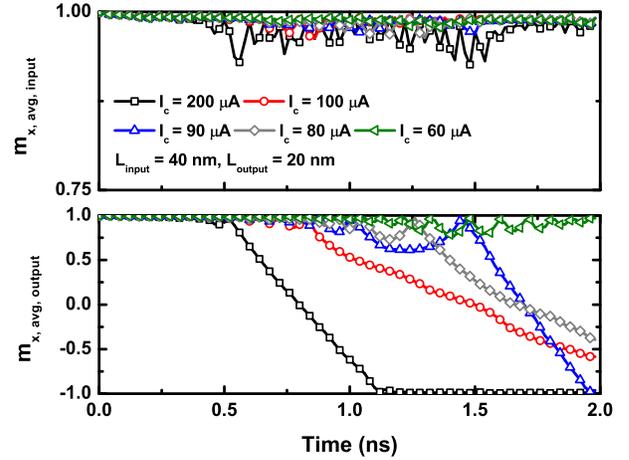}
\caption{Time evolution of average magnetizations of FM wires for the input (top) and the output (bottom) in a SV inverter under different magnitudes of driving current.}
\label{fig7}
\end{figure}

In addition to the damping process and contact area, the device non-reciprocity can also be improved by smaller driving currents as shown in Fig. \ref{fig7}, where the input's magnetization becomes less disturbed as the applied current is reduced. However, the speed of DW creation becomes slower at smaller driving current. This is because STT at the output is significantly reduced when both input and output current densities are lower. Note that a DW cannot be created because of insufficient STT if the driving current is too small (e.g. olive in Fig. \ref{fig7}).  

\subsection{Buffer, Inverter, 3-input Majority Gate, and DW/Spin current Transducer}
\begin{figure}[htp]
\includegraphics[width = 3.5in]{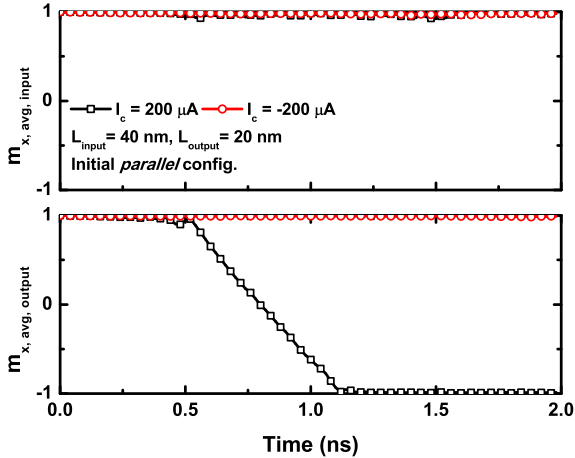}
\caption{Time evolution of average magnetizations of FM wires for the input (top) and the output (bottom) in a SV device with initial parallel alignment under positive and negative driving currents.}
\label{fig8}
\end{figure}

\begin{figure}[htp]
\includegraphics[width = 3.5in]{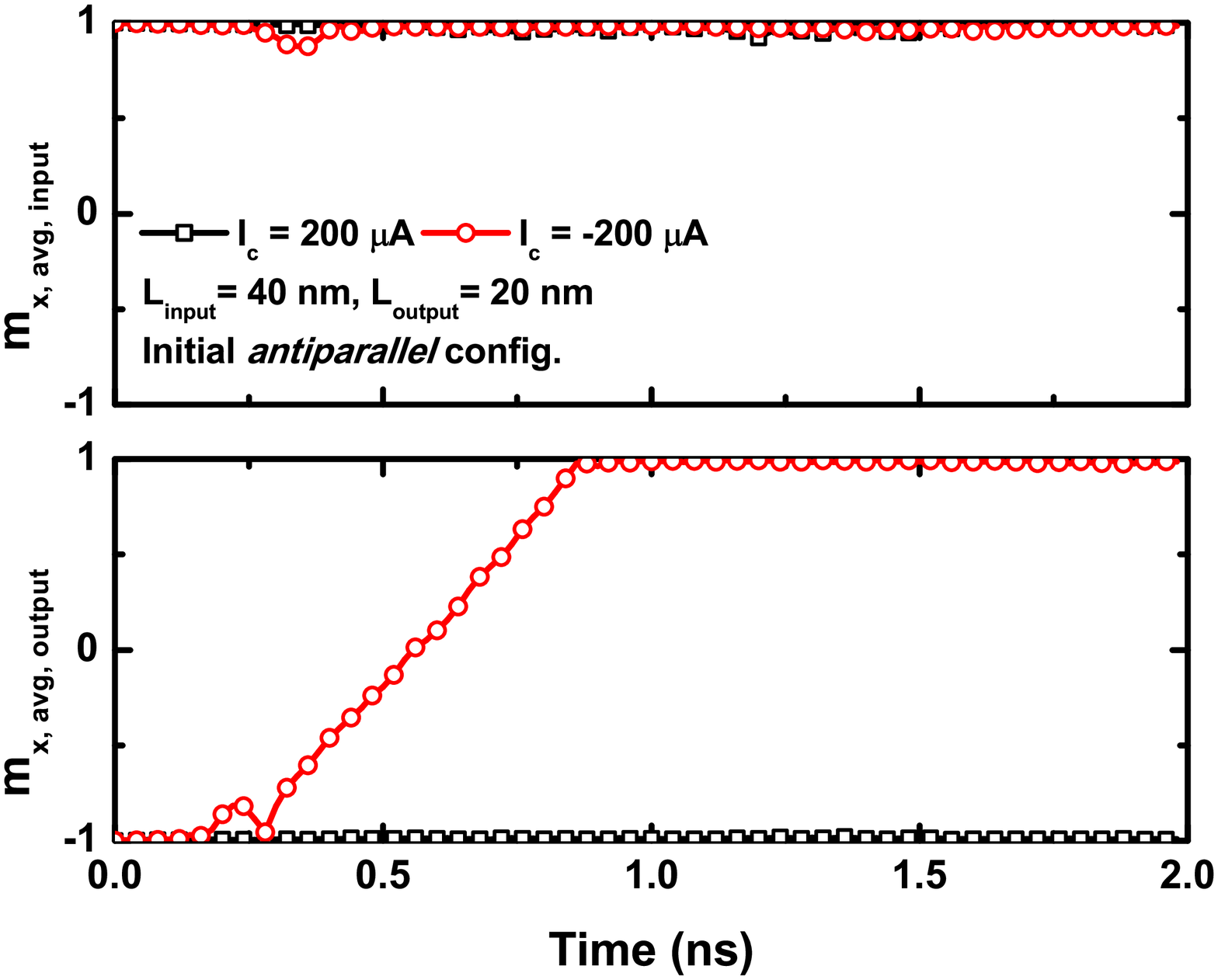}
\caption{Time evolution of average magnetizations of FM wires for the input (top) and the output (bottom) in a SV device with initial anti-parallel alignment under positive and negative driving currents.}
\label{fig9}
\end{figure}

\begin{figure}[htp]
\includegraphics[width = 3.5in]{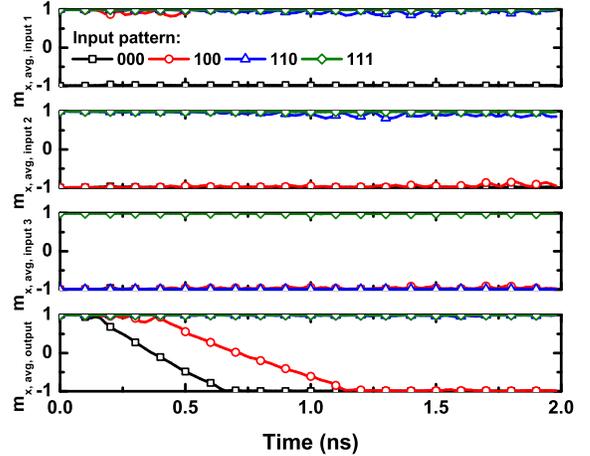}
\caption{Time evolution of average magnetizations of FM wires in a 3-input majority gate based on SVs for the three inputs and the output (bottom). $-100\mu$A is applied at each input. The input and output contact lengths are $40$ and $20$nm, respectively.}
\label{fig10}
\end{figure}

As explained in Section II, the proposed device prefers parallel and anti-parallel configurations under negative and positive driving currents, respectively, and a buffer as well as an inverter can be implemented based on that. Figs. \ref{fig8} and \ref{fig9} show that by properly sizing contact areas and using reasonable damping coefficient at the end ($\alpha=0.18$) for the non-reciprocity, the device can act as an inverter (a buffer) as positive (negative) current is applied. Note that DW creation from the anti-parallel state is faster than that from the parallel one, which is consistent with a substantial asymmetry between the differential torque near parallel and anti-parallel alignment, predicted by all semiclassical calculations of transport and STT in metallic multilayers \cite{Brataas2006157,Stiles2006}. Fig. \ref{fig10} also demonstrates that a 3-input majority gate can be operated normally under the proposed scheme. Note that the switching responses are different for the input patterns being 000 and 100 since net STT at the output are different; thus, clocking a circuit with majority gates may be non-trivial.

In addition to logic gates, the proposed scheme can also be used as a transducer between the DW and spin current. As a result, a hybrid interconnect system combing the advantages of DW automotion and spin diffusion can be constructed to propagate spin information. For instance, the interconnect using automotion is energy-free, but it is difficult to bend a DW interconnect to have a $90^{\circ}$ turn due to pinning sites at the corners. However, spin-diffusive interconnects can still work well at abrupt turning angles. Therefore, the wiring in spin circuits can become more flexible when this hybrid scheme is applied.

\subsection{Comparison to NLSVs and CMOS Circuits}
To explore the real potential of the proposed scheme, it is required to compare the scheme based on NLSVs, which can also provide a complete set of Boolean functions \cite{7130566}. However, to have a fair comparison between two schemes, here a performance optimization of NLSVs by engineering the contact areas is briefly discussed. The equivalent spin circuit for a single NLSV device can be found in Supplementary Materials. Fig. \ref{fig11} shows that as the output contact length increases, a DW is created slower with fixed injected spins at the input. This is because in NLSVs, STT at the output only depends on injected spins at the input; therefore, a larger FM region needs longer time to be switched under the same amount of STT. If the input contact length is increased, meaning that both input current density and injected spins are reduced, STT at the output is weakened and may not be strong enough to create a DW as shown in Fig. \ref{fig12}. As a result, based on Figs. \ref{fig11} and \ref{fig12}, smaller input and output contact lengths are desired in NLSVs to provide strong STT exerted on the output. With the optimized contact size, Fig. \ref{fig13} investigates the minimum driving current in NLSVs, and it is found that the minimum driving current in both schemes are similar (see Fig. \ref{fig7}); however, the current density in SVs is much smaller than that in NLSVs. This is mainly because in NLSVs, a significant part of injected spins are directly flowing into the ground, rather than contributing STT at the output. In SVs, all the spins participate the switching process. Note that the non-reciprocity in NLSVs is realized by the asymmetry of the non-local structure, and sizing different input and output contact areas is only for performance optimization, which is very different from SVs.

\begin{figure}[htp]
\includegraphics[width = 3.5in]{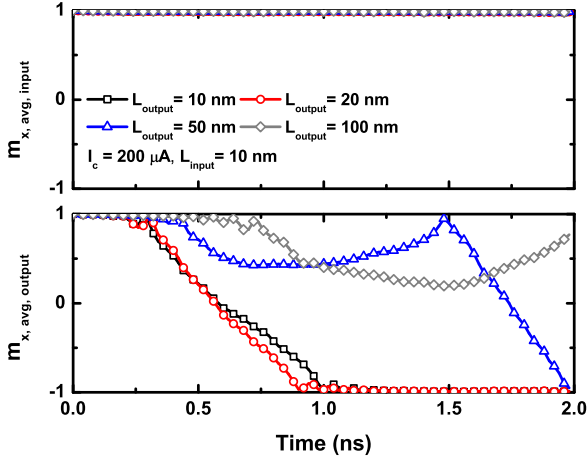}
\caption{Time evolution of average magnetizations of FM wires for the input (top) and the output (bottom) in a NLSV inverter with different output contact lengths.}
\label{fig11}
\end{figure}

\begin{figure}[htp]
\includegraphics[width = 3.5in]{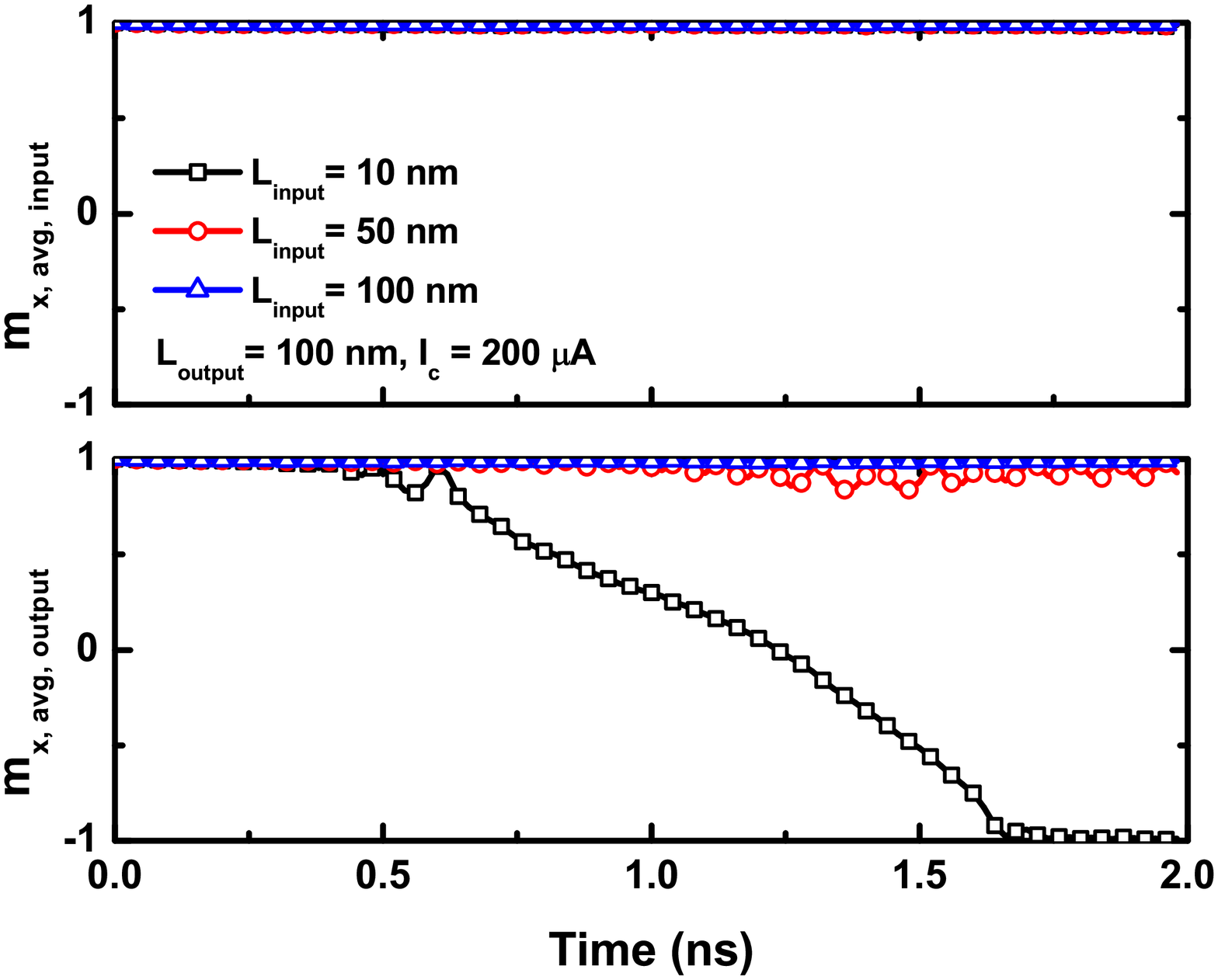}
\caption{Time evolution of average magnetizations of FM wires for the input (top) and the output (bottom) in a NLSV inverter with different input contact lengths.}
\label{fig12}
\end{figure}

\begin{figure}[htp]
\includegraphics[width = 3.5in]{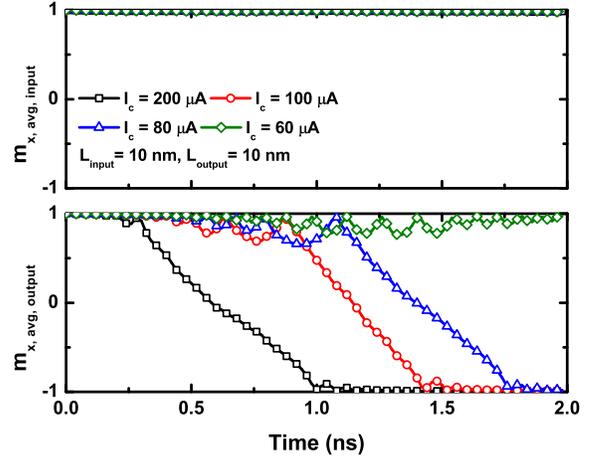}
\caption{Time evolution of average magnetizations of FM wires for the input (top) and the output (bottom) in a NLSV inverter under different magnitudes of driving current.}
\label{fig13}
\end{figure}

It is well known that the energy dissipation of current-driven STT logic devices is mainly due to the wiring network linked to the global power supply rather than the device itself \cite{7076743}. Here $300\Omega$ is assumed as the wiring resistance from the device to the global power supply for the contact length equal to $10$nm, and the switching energy is calculated using $E=I^{2}R\tau$, where $E$ is the switching energy, $I$ is the driving current, $R$ is the wiring resistance, and $\tau$ is the current pulse duration. Using $I=200\mu$A as an example, the corresponding energy for a single switching in a NLSV device is $18$fJ (black in Fig. \ref{fig13}). Similarly, with the same driving current, a SV device only dissipates $6.7$fJ due to lower resistance of the wiring network (black in Fig. \ref{fig7}), and thus provides a new low-power option for the STT-driven logic family. Since the proposed scheme can be operated at lower current density, electromigration induced by large current density in Cu-based NLSVs \cite{Su2015} can be significantly mitigated. Table. \ref{tab2} summarizes the comparison between schemes based on SVs and NLSVs.

\begin{table}[ht!]
\centering
\caption{A performance comparison between schemes based on SVs and NLSVs. $200\mu$A is applied to both structures to simulate inverters, and the shunt path in the NLSV is $30$nm. The delay ($\tau$) is defined as the total time required for DW creation in the beginning of the FM wire and DW automotion to the end of the FM wire. $E$ is the switching energy.}
\renewcommand{\arraystretch}{1.2}
\begin{tabular}{c c c c}
\hline
\hline
($L_{input}$, $L_{output}$) & $I_{c}$ & $E$ & $\tau$ \\
\hline
SV ($40$nm, $20$nm) & $200\mu$A & $6.7$fJ & $1.12$ns \\
NLSV ($10$nm, $10$nm) & $200\mu$A & $18$fJ & $1$ns \\ 
\hline
\hline
\end{tabular}
\label{tab2}
\end{table}

In addition to the non-volatility and the eliminating of the static power in circuits, another major advantage of the proposed scheme is a straightforward implementation of a majority gate as shown in Fig. \ref{fig2}, which largely enables a significant reduction in the required circuit layout area. Table. \ref{tab3} shows a performance comparison for a 3-input majority gate based on the proposed devices (Fig. \ref{fig10}) and low-power CMOS transistors (15nm CMOS technology node) \cite{7076743}. From Table. \ref{tab3}, it is shown that even though the switching energy (or dynamic power) using the proposed scheme is still much higher than that using the CMOS counterpart due to a slow magnetic response of the FM metal to STT, the circuit area under the proposed scheme is significantly reduced.

\begin{table}[ht!]
\centering
\caption{A performance comparison for a majority gate using the proposed devices (Fig. \ref{fig10}) and the low-power (LP) CMOS switches (\cite{7076743} F=15nm, 2018 technology node in the 2013 edition of ITRS). $E$, $\tau$, and $Area$ are the switching energy, critical delay, and required circuit area, respectively. The spacing between FM interconnects is assumed to be $20$nm. A $3$-input majority gate function is given as $O=AB+BC+CA$, where $A$, $B$, $C$ are the inputs and $O$ is the output. In the CMOS implementation, a $3$-input majority gate is composed of three 2-input NAND and one 3-input NAND gates, which require at least 18 CMOS digital switches with routing interconnects in different metal layers to minimize the area.}
\renewcommand{\arraystretch}{1.2}
\begin{tabular}{c c c c}
\hline
\hline
 & $E$ & $\tau$ & $Area$ \\
\hline
This work & $20.1$fJ & $1.14$ns & $0.01\mu$m$^{2}$ \\
LP CMOS \cite{7076743} & $0.069$fJ & $0.042$ns & $0.166\mu$m$^{2}$ \\ 
\hline
\hline
\end{tabular}
\label{tab3}
\end{table}
\section{Conclusion}
This paper presents a novel scheme using SVs and FM wires to perform digital computation, and justifies the proposed concept through comprehensive simulations including spin transport in metallic multilayers and stochastic magnetization dynamics. The proposed scheme offers a new option to implement low-power logic using current-driven STT due to removing the shunt path in NLSVs, and is more suitable in the path of scaling because of using FM wires to store bits, rather than single-domain FM metals. Furthermore, the proposed concept can also be viewed as a transducer between spin current and magnetic DW, which may significantly increase the flexibility in the wiring network of spin interconnects.

\bibliographystyle{IEEEtran}
\bibliography{IEEEabrv,ieee_proc}

\end{document}